\pdfoutput=1
\documentclass[journal,twoside,web]{ieeecolor}
\usepackage{tmi}
\usepackage{cite}
\usepackage{amsmath,amssymb,amsfonts}
\usepackage{algorithmic}
\usepackage{graphicx}
\usepackage{textcomp}
\usepackage{multirow}
\usepackage{array}
\def\BibTeX{{\rm B\kern-.05em{\sc i\kern-.025em b}\kern-.08em
    T\kern-.1667em\lower.7ex\hbox{E}\kern-.125emX}}
\begin{document}
\bstctlcite{IEEEexample:BSTcontrol}
\title{Deep Learning-based Synthetic High-Resolution In-Depth Imaging Using an Attachable Dual-element Endoscopic Ultrasound Probe}
\author{Hah Min Lew,~Jae Seong Kim,~Moon Hwan Lee,~Jaegeun Park,~Sangyeon Youn,~Hee Man Kim,~Jihun Kim,~and Jae Youn Hwang
\thanks{Manusript received xxx; revised xxx. 
This work was partly supported by the Korea Medical Device Development Fund Grant funded by the Korean government (Ministry of Science and ICT, Ministry of Trade, Industry and Energy, Ministry of Health \& Welfare, the Ministry of Food and Drug Safety) (RS-2020-KD000125, 9991006798) and partly supported by the Korea Medical Device Development Fund Grant funded by the Korean government (Ministry of Science and ICT, Ministry of Trade, Industry and Energy, Ministry of Health \& Welfare, the Ministry of Food and Drug Safety) (RS-2022-00141185). 
\textit{(Hah Min Lew and Jae Seong Kim contributed equally to this work.)} 
\textit{(Corresponding author: Jae Youn Hwang.)}}
\thanks{Hah Min Lew was with the
Department of Electrical Engineering and Computer Science, DGIST, Daegu 42988, South Korea. 
He is now with the Deep Learning Research Team, KLleon R\&D Center, Seoul 04637, South Korea (e-mail: hahmin.lew@klleon.io).}
\thanks{Jae Seong Kim was with the Department of Information and Communication Engineering, DGIST, Daegu 42988, South Korea.
He is now with the Software Team, ALPINION MEDICAL SYSTEMS, Anyang-si 14117, South Korea (e-mail: jaeseong.kim@alpinion.com).}
\thanks{Jaegeun Park, Moon Hwan Lee, and Sangyeon Youn are with the
Department of Electrical Engineering and Computer Science, DGIST, Daegu 42988, South Korea
(e-mail: nick123478@dgist.ac.kr; moon2019@dgist.ac.kr; ttorxp12@dgist.ac.kr).}
\thanks{Hee Man Kim is with the
Department of Health Promotion Center, Yonsei University College of Medicine, Gangnam Severance Hospital, Seoul 06273, South Korea (e-mail: doctorman@yonsei.ac.kr).}
\thanks{Jae Youn Hwang is with the
Department of Electrical Engineering and Computer Science and The Interdisciplinary Studies of Artificial Intelligence, DGIST, Daegu 42988, South Korea (e-mail: jyhwang@dgist.ac.kr).}}

\maketitle

\begin{abstract}
Endoscopic ultrasound (EUS) imaging has a trade-off between resolution and penetration depth. 
By considering the in-vivo characteristics of human organs, it is necessary to provide clinicians with appropriate hardware specifications for precise diagnosis. 
Recently, super-resolution (SR) ultrasound imaging studies, including the SR task in deep learning fields, have been reported for enhancing ultrasound images. 
However, most of those studies did not consider ultrasound imaging natures but rather they were conventional SR techniques based on downsampling of ultrasound images. 
In this study, we propose a novel deep learning-based high-resolution in-depth imaging probe capable of offering low- and high- frequency ultrasound image pairs.
We developed an attachable dual-element EUS probe with customized low- and high-frequency ultrasound transducers under small hardware constraints.
We also designed a special geared structure to enable the same image plane. 
The proposed system was evaluated with a wire phantom and a tissue-mimicking phantom.
After the evaluation, 442 ultrasound image pairs from the tissue-mimicking phantom 
were acquired. 
We then applied several deep learning models to obtain synthetic high-resolution in-depth images, thus demonstrating the feasibility of our approach for clinical unmet needs.
Furthermore, we quantitatively and qualitatively analyzed the results to find a suitable deep-learning model for our task. 
The obtained results demonstrate that our proposed dual-element EUS probe with an image-to-image translation network has the potential to provide synthetic high-frequency ultrasound images deep inside tissues.
\end{abstract}


\section{Introduction}
\label{sec:introduction}
\IEEEPARstart{E}{ndoscopic} ultrasound (EUS) is a useful modality that provides anatomical information of the gastrointestinal (GI) tract at the submucosal membrane for diagnosis of various human diseases related to the GI tract ~\cite{byrne2002gastrointestinal, puli2008staging}.
Unlike abdominal ultrasound imaging, EUS can provide better visualization of the GI tract since it is less affected by gas and/or fat interposition~\cite{almansa2011role}.
EUS devices, first reported in the early 1980s, have been developed into various EUS imaging modalities to provide abundant clinical information combined with other imaging techniques or by upgrading the design of EUS probes~\cite{dimagno1980ultrasonic, strohm1980ultrasonic, ji2015intravascular, wang2017development, lay2018design, qiu2020ultrasound, kim2020forward}.

Generally, the frequency range for medical applications is 2 to 15 MHz~\cite{jensen2007medical}.
Several clinical observations reported that the appropriate frequency range of ultrasound transducers should be at least 5 MHz to distinguish beyond the GI wall, ~\cite{nylund2017efsumb, aibe1986fundamental, boscaini1986transrectal, kimmey1989histologic}.
On the other hand, for the valid detection of intestinal or esophageal layers and related diseases, a high-frequency range of around 10 to 20 MHz should be selected~\cite{nylund2017efsumb}.
A proper frequency range should be selected because ultrasound imaging has a trade-off between penetration depth and resolution~\cite{szabo2004diagnostic, shung2005diagnostic}.
Naturally, clinicians demand the highest resolution image while maintaining the proper imaging depth to acquire abundant clinical information for better diagnosis. It is accordingly necessary to provide a high-resolution image with sufficient clinical information deep inside tissues for various organs using EUS probes.

\begin{figure*}[t]
\centerline{\includegraphics[width=2\columnwidth]{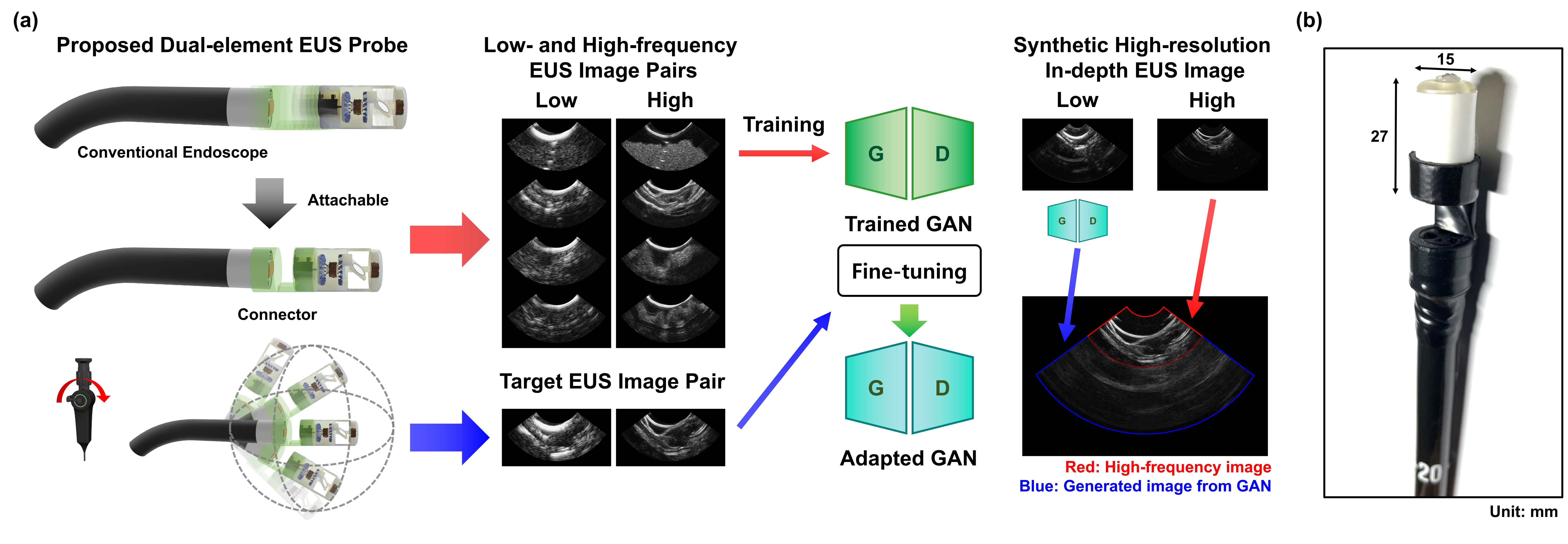}}
\caption{(a) Illustration for the usage of the proposed dual-element EUS probe with fine-tuning of a GAN trained on a paired dataset. (b) A photograph of the assembled dual-element EUS probe into a conventional endoscope.}
\label{fig:0}
\end{figure*}

To enhance the performance of ultrasound imaging, several approaches have been studied. 
By using microbubbles with advanced processing techniques, super-resolution (SR) ultrasound imaging is utilized~\cite{oreilly2013super, errico2015ultrafast, couture2018ultrasound, christensen2020super, kim2021compressed, kim2022improved}. 
However, these methods require an injection of additional contrast agents. In addition, since these techniques compound various frames according to the blinking of microbubbles to construct one high-resolution image~\cite{kim2021compressed}, they are not suitable for endoscopic applications.
Some researchers have suggested that low-resolution ultrasound images can be enhanced by deep learning-based approaches. The SR task in the deep learning area is to find missing pixels in a low-resolution image when it is scaled up to obtain a high-resolution image~\cite{yang2019deep}.
The authors of~\cite{temiz2020super} proposed a novel convolutional neural network to construct ultrasound high-resolution B-mode images from downsampled low-resolution images.
With a similar training scheme for producing high-resolution images, Liu et al.~\cite{liu2021perception} utilized a new generative adversarial network (GAN) with a perception cycle consistency loss.
Both approaches, however, are not suitable for the purpose of constructing high-resolution ultrasound images since ultrasound imaging is based on bulk mechanical properties according to how a point spread function (PSF) of ultrasound transducers interacts with objects~\cite{szabo2004diagnostic, shung2005diagnostic}.
If the original ultrasound image is downsampled to form a high-resolution image, the deep learning model learns to reconstruct the original ultrasound image from the downsampled image rather than construct an actual high-resolution ultrasound image based on the proper PSF of high-frequency ultrasound.
Therefore, it is necessary to prepare paired images from low- and high-frequency ultrasound for an appropriate training scheme for high-resolution image generation.
Recently, attempts to prepare paired ultrasound images for generating high-resolution images based on GANs have been reported.
Wang et al.~\cite{wang2019high} proposed a new sparse skip connection U-Net for high-resolution image reconstruction by using simulation data (3 MHz, 5 MHz, and 8 MHz) from Field II program~\cite{jensen1996field} and various ultrasound images (6 MHz, 7 MHz, and 7.5 MHz) from several ultrasound imaging devices.
The authors of~\cite{goudarzi2019multi} proposed a new training scheme that can achieve dynamic focusing for ultrasound images without any trade-off between the frame rate and resolution.
Asiedu et al.~\cite{asiedu2022generative} presented image quality enhancement according to the transmit voltage, which can highly determine the signal-to-noise ratio and contrast-to-noise ratio of ultrasound images.
However, there are still no studies focused on translating low-frequency images to high-frequency images because without an optimized hardware system that can obtain the same image plane from a wide frequency range, such as 5 MHz to 20 MHz.
In this study, we firstly propose a deep learning-based dual-element EUS probe for synthetic high-resolution in-depth imaging to overcome the limitations of conventional ultrasound imaging properties.
The proposed system can obtain paired images providing complementary information through a novel geared-structure design that can have the same imaging plane.
We fabricated and tested an attachable dual-element EUS probe similar to an ultrasound capsule endoscope, which has the highest hardware constraints among EUS techniques~\cite{wang2017development,qiu2020ultrasound}.
To secure sufficient reliability from a clinical point of view, we fabricated customized low- (5.1 MHz) and high-frequency (18.3 MHz) transducer pairs.
The low-frequency ultrasound transducer is utilized for obtaining in-depth information, while the high-frequency ultrasound transducer is utilized to obtain high-resolved spatial information.
In addition to the complementary EUS imaging, we also analyze the feasibility of providing high-resolved in-depth information through a deep learning model for preliminary study.
Various GANs were utilized for pseudo-high-frequency imaging because synthetic high-resolution ultrasound imaging is similar to the image-to-image translation task between images from the low- and high-frequency ultrasound transducers.
In addition, we applied a representative network of the SR task in the deep learning area to analyze the feasibility of the SR task for ultrasound imaging.
We also demonstrated synthetic high-resolution in-depth EUS imaging with fine-tuning of a GAN trained on a paired dataset.
Fig.~\ref{fig:0} depicts the concept of synthetic high-resolution in-depth EUS imaging using the proposed dual-element EUS probe.
Finally, we compare the associated qualitative and quantitative results of deep learning experiments using low- and high-frequency ultrasound imaging.
We thus demonstrate the potential of pseudo-high-frequency in-depth imaging using paired ultrasound images.
\section{MATERIALS AND METHODS}
\subsection{Configuration of a Dual-element EUS Probe}
\begin{figure}[b]
\centerline{\includegraphics[width=\columnwidth]{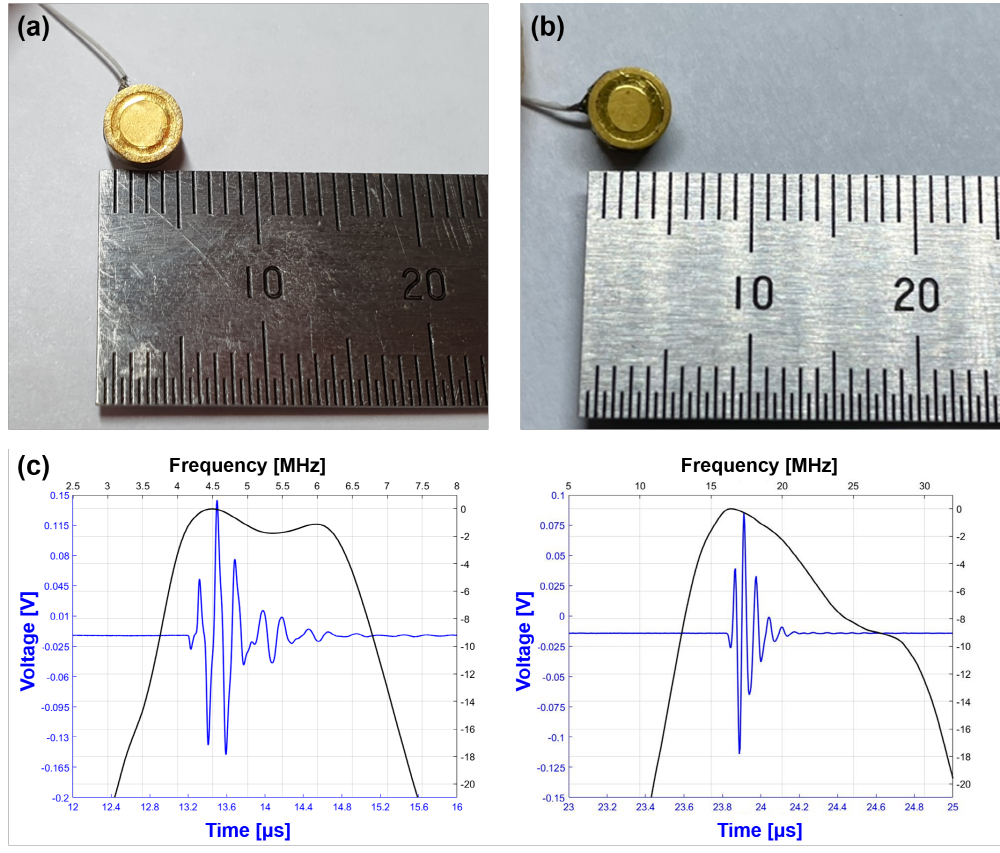}}
\caption{Characteristics of fabricated ultrasound transducers (a) A low-frequency ultrasound transducer with a center frequency of 5.1 MHz and 52 \% of -6 dB bandwidth. (b) A high-frequency ultrasound transducer with a center frequency of 18.3 MHz and 51 \% of -6 dB bandwidth. (c) Pulse-echo characteristics of the ultrasound transducers.}
\label{fig:1}
\end{figure}

\begin{figure}[t]
\centerline{\includegraphics[width=\columnwidth]{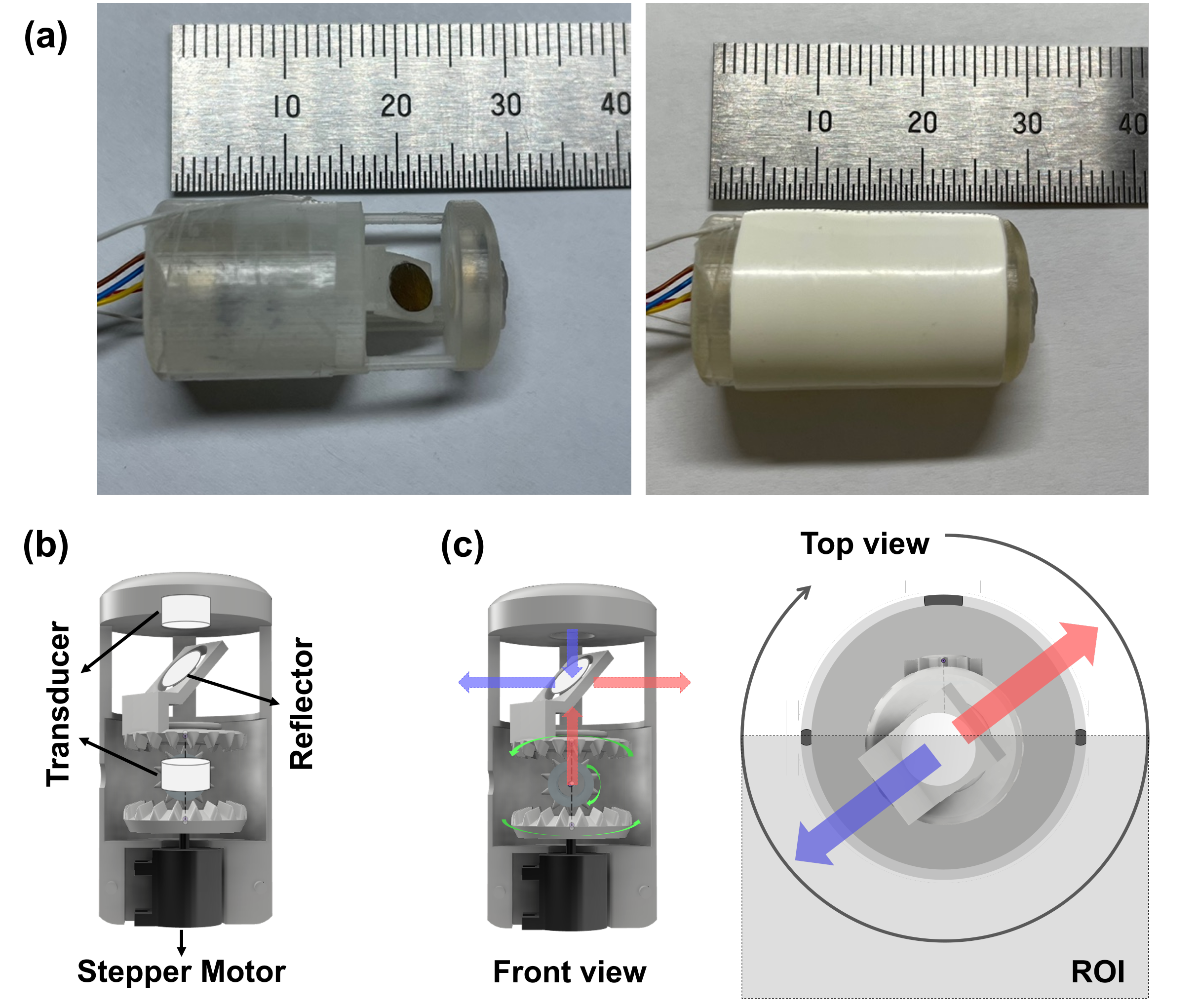}}
\caption{(a) The structure of the proposed EUS probe with or without an acoustic window. (b) Positions of each component. (c) Illustration of scanning diagrams with estimated beam paths.}
\label{fig:2}
\end{figure}

Two different types of compact ultrasound transducers are required for the proposed probe. 
Therefore, 18.3 MHz and 5.1 MHz customized ultrasound transducers were fabricated for high-resolution in-depth imaging. 
The developed low- and high-frequency ultrasound transducers are shown in Fig.~\ref{fig:1}(a) and (b), respectively.
The low-frequency ultrasound transducer is fabricated with lead zirconate titanate (PZT-5H), and the high-frequency ultrasound transducer is made of lithium niobate (LiNbO3). 
2 – 3 $\mu$m silver particles are used for the first matching layer. 
Also, a parylene coating is used for the second matching layer of the high-frequency ultrasound transducer. 
E-solder 3022 (Von Roll) is used for the backing layer.
The thickness of each layer is determined by modeling software (PiezoCAD, Sonic concepts). 
The pulse-echo characteristics of the customized ultrasound transducers are shown in Fig.~\ref{fig:1}(c).

The structure of the proposed EUS probe is shown in Fig.~\ref{fig:2}. 
The EUS probe mainly comprises a motor and gearbox for adjusting the reflection and irradiation direction for both ultrasound transducers.
To increase the frame rate by rotating in one direction while preserving the high scanning speed and avoiding tangling of the co-axial cable, the structure utilizes a customized geared stainless reflector inclined at an angle of 45°.
Integrating a gearbox with the motor shaft and mounting an ultrasound reflector on the gearbox mechanically transfers torque to the reflector. 
Therefore, the irradiation direction of the ultrasound beam is available for high-speed rotation without tangling the coaxial cables of the ultrasound transducers located on both sides of the reflector throughout the full range of motion.

The 5.1 MHz ultrasound transducer is mounted on the top of the capsule, and the 18.3 MHz ultrasound transducer is mounted below the reflector. 
Therefore, when two different frequencies of the ultrasound beams are irradiated toward the reflector mounted on the gearbox, the ultrasound beams are reflected in a perpendicular direction to the circumference of the probe with a phase difference of 180° while scanning the same image plane.

The EUS probe has three pillars at the scanning plane to support the structure and indicate the region of interest (ROI). 
The pillars are located 7 mm from the center of the probe, and the two thin pillars are aligned at 90° to the thick pillar. 
The ROI is located at 180° of the image plane, where pillars do not exist.

Finally, a PEBAX 4033-based sheath for an acoustic window is applied to the outside of the probe to transmit and receive the ultrasound waves.
A PEBAX 4033 has a slight difference in acoustic impedance between the water and tissue. 
The thickness of the acoustic window is 500 $\mu$m. 
Therefore, the total diameter of the proposed EUS probe is 15 mm, and the height is 27 mm.

\begin{figure}[t]
\centerline{\includegraphics[width=\columnwidth]{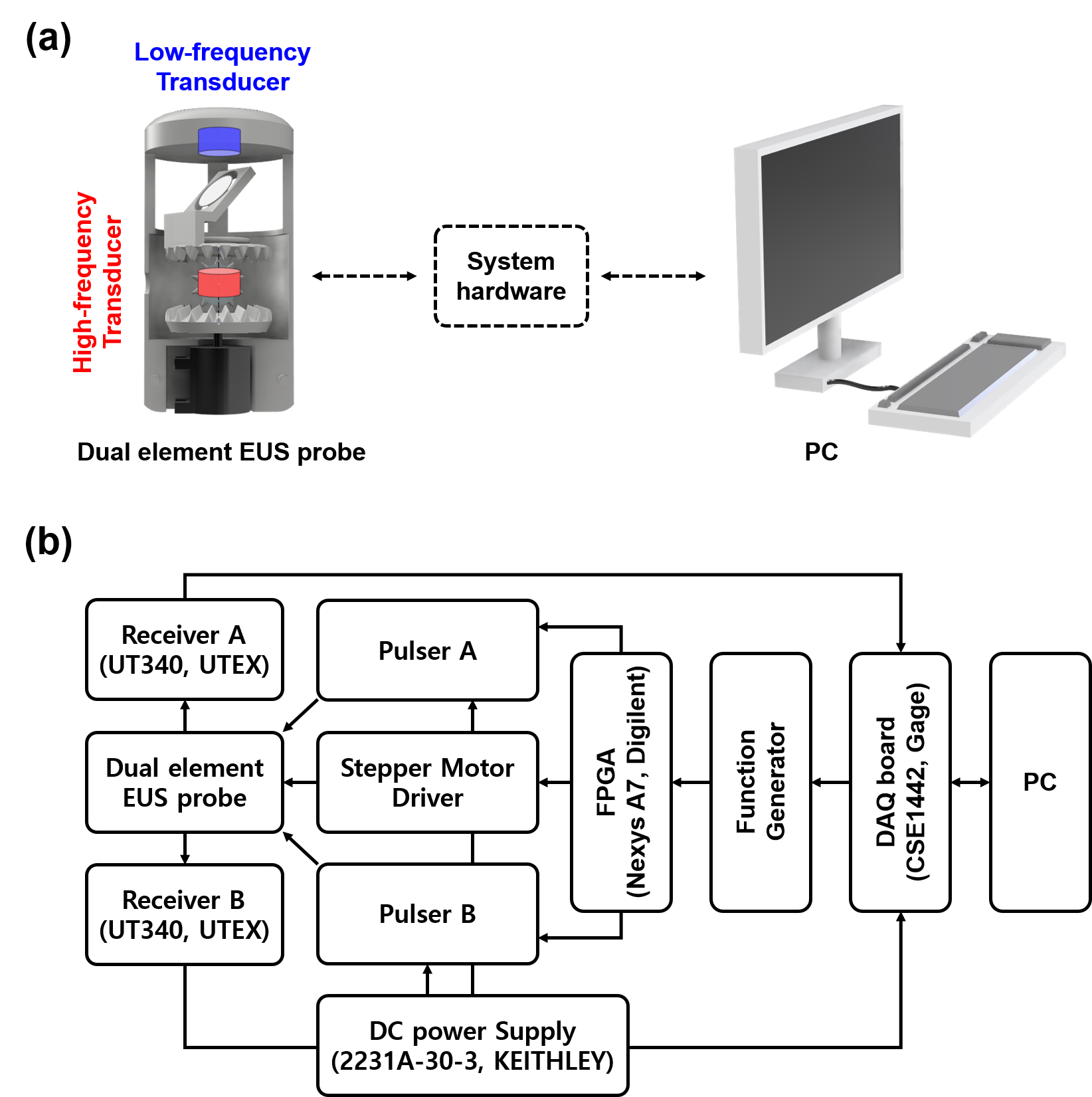}}
\caption{Experimental setup for the proposed system with the dual-element probe. (a) Overall schematics of the system. (b) System block diagram.}
\label{fig:3}
\end{figure}

The hardware schematic of the proposed system with the dual-element probe is shown in Fig.~\ref{fig:3}. 
The proposed system mainly comprises a field-programmable gate array (FPGA) board used for controlling the entire system, ultrasound pulser/receiver pairs, and the EUS probe.
A multi-channel data acquisition (DAQ) board (CSE1442-4GS, Gage Applied Technologies) is used for triggering and recording ultrasound signals.
A function generator (AFG3252C, Tektronics) controls the FPGA board (Nexys A7, Digilent) according to the pulse repetition frequency (PRF).
The ultrasound pulser pairs consist of a T/R switch, MOSFET driver (MD1213, Microchip Technology Inc.), and MOSFET pair (TC6320).
The pulser boards generate a single bipolar pulse of 30 peak-to-peak voltage.
For receiving ultrasound echo signals, receiver pairs (UT 340, UTEX) are used.
To scan the image plane, the FPGA board transmits the trigger to the stepper motor driver (STSPIN220, STMicroelectronics), which drives the stepper motor (SMS6-F40, MinebeaMitsumi Inc) via micro-stepping to rotate a specially designed geared reflector through 360°.
Because the gearbox rotates at 1293.1 rpm, the frame rate of the system is 21.6 fps.

\subsection{System Evaluation}

\begin{figure}[t]
\centerline{\includegraphics[width=\columnwidth]{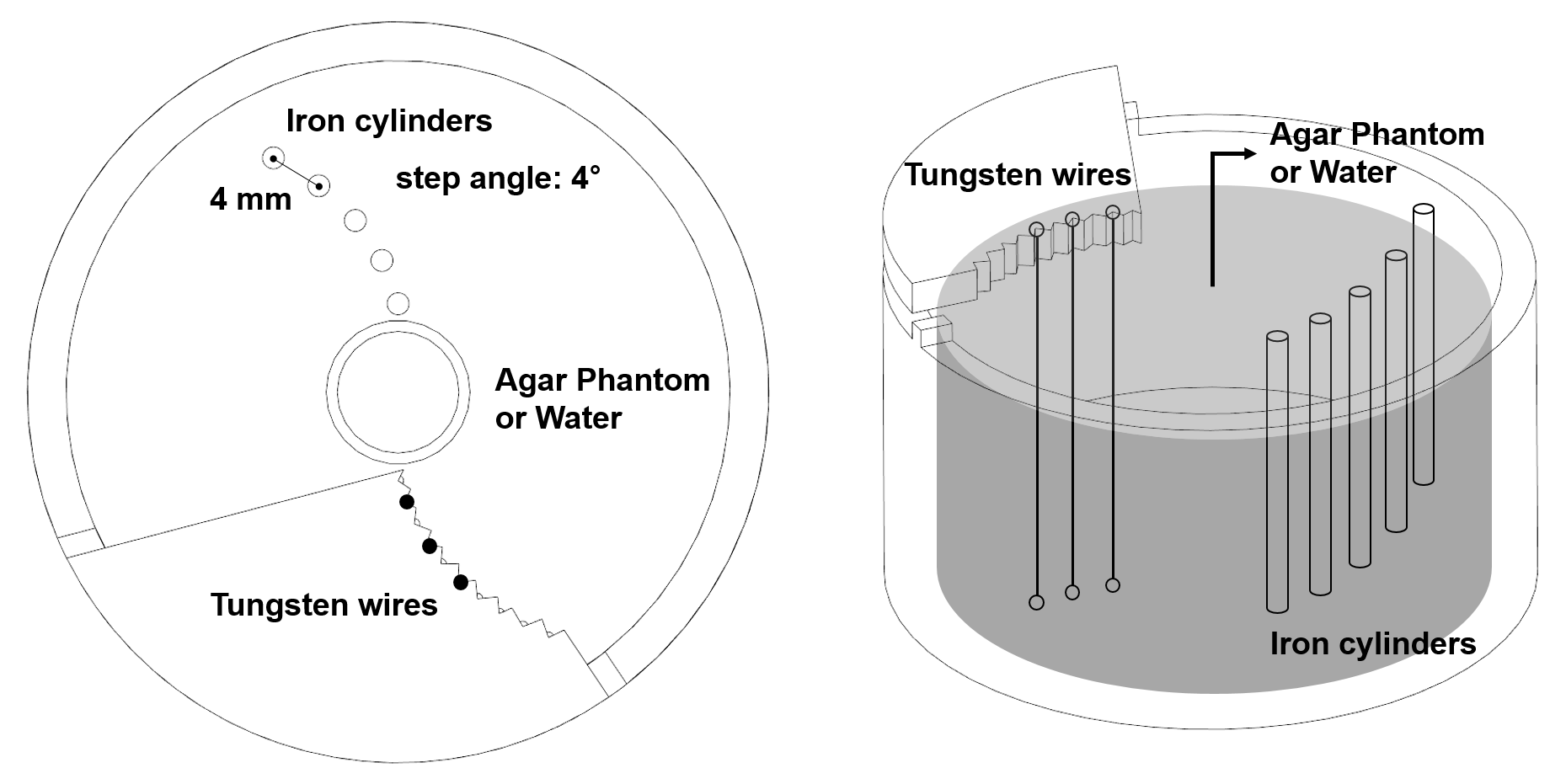}}
\caption{The phantom structure for evaluation of the proposed system.}
\label{fig:4}
\end{figure}

To demonstrate the feasibility and evaluate the performance of the proposed dual-element EUS system, rotational phantoms are constructed, and the detailed structures are shown in Fig.~\ref{fig:4}. 
The rotational phantoms are designed to place the proposed system at the center. 
The diameters of wire targets and pillar targets are 100 $\mu$m and 4 mm, respectively.
To demonstrate the imaging performance in a biological environment, a tissue-mimicking phantom is utilized to evaluate the proposed system. 
The agar-based phantom is made with graphite powder and n-propanol to model the echogenicity and speed of sound in human tissue, respectively.

To evaluate the performance of the proposed system, we calculate the spatial resolution, echo signal-to-noise ratio (eSNR), contrast-to-noise ratio (CNR), and speckle’s signal-to-noise ratio (SSNR) of ultrasound images.

To evaluate the spatial resolution, the lateral and axial resolution of the second wire target at the underwater phantom are calculated based on the full width at half maximum (FWHM) criterion.
Note that the spatial resolution is evaluated based on the target located near the focal depth of the high-frequency ultrasound transducer.

The eSNR is calculated to evaluate the penetration depth of the proposed system~\cite{park2013pulse}. 
To evaluate the eSNR of ultrasound images, we crop the areas of the wire targets over the depth and homogeneous area of the tissue-mimicking phantom. 
Because water is homogeneous, the area filled with water can be considered an area containing noise information.

\begin{equation}eSNR=20\log_{10} (\frac{\textrm{max}(signal)}{\sigma_{\textrm{noise}}}),\label{eq:1}\end{equation}

where $\textrm{max}(signal)$ is the maximum amplitude value of ultrasound echo signals and $\sigma_{\textrm{noise}}$ is the standard deviation of noises.

For a quantitative assessment of resolution, we use metrics of CNR and SSNR.
The related equations are shown in (\ref{eq:2}) and (\ref{eq:3}).

\begin{equation}CNR=\frac{|\mu_{T}-\mu_{B}|}{\sqrt{{\sigma_{T}}^2+{\sigma_{B}}^2}},\label{eq:2}\end{equation}

where $\mu_{T}$ and $\mu_{B}$ represent the average magnitude of the B-mode image in the homogeneous and background regions in the tissue-mimicking phantom, $\sigma_{T}$ and $\sigma_{B}$ are the standard deviations of the B-mode image in the homogeneous and background regions, respectively.

\begin{equation}SSNR=\frac{\mu}{\sigma},\label{eq:3}\end{equation}

where $\mu$ and $\sigma$ represent the mean intensity and standard deviation of the extracted speckle region, respectively.

\subsection{Preparation of a Paired Image Dataset}

\begin{figure}[t]
\centering
\includegraphics[width=\columnwidth]{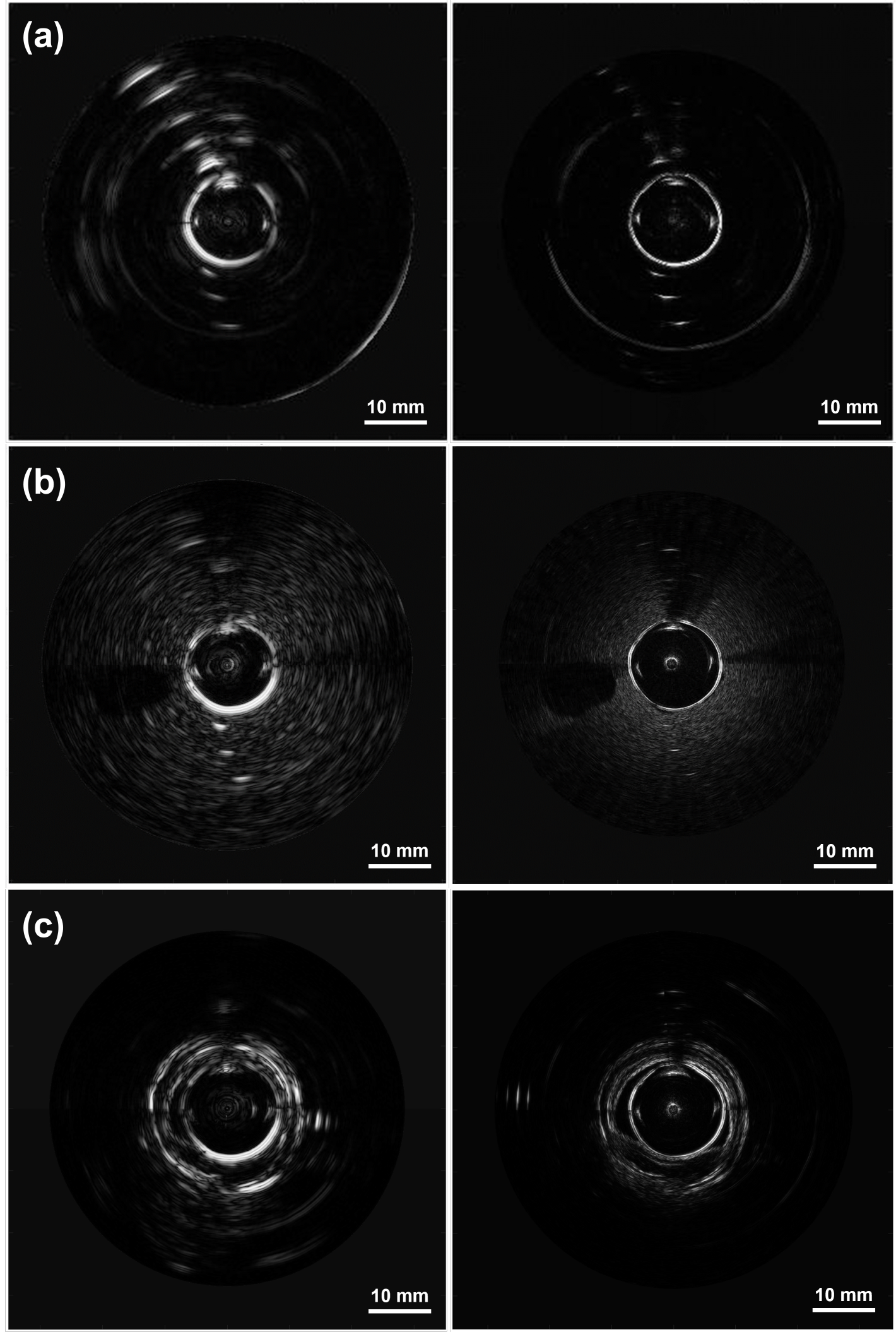}
\caption{Paired images were obtained from the proposed EUS probe. Left figures represent images from the low-frequency ultrasound transducer and right figures represent images from the high-frequency ultrasound transducer (a) A wire phantom. (b) A tissue-mimicking phantom. (c) A porcine esophagus.}
\label{fig:5}
\end{figure}

By using the proposed EUS probe, we obtained 442 images of low- and high-frequency pairs, as shown in Fig.~\ref{fig:5}.
Low- and high-frequency ultrasound image pairs were obtained from tissue-mimicking phantom and porcine organs in a GI tract such as an esophagus, stomach, and intestine.
The lateral range was determined by the number of scanlines at the ROI.
The depth of the input data was determined by the focal depth of the high-frequency ultrasound transducer. 
The depth is 2 cm, and the lateral range is 106°, which represents the ROI of the proposed EUS probe.
Thus, the image size was 436 × 1000 pixels in width and height.

\subsection{Deep learning Experimental Setup}

In the deep learning experiments, seven different GANs were trained and tested using the paired dataset, including 109 pairs of phantoms and 333 pairs of porcine organs in the GI tract.
We divided the dataset into five groups for 5-fold cross-validation (90, 88, 88, 88, and 88 image pairs). 
The input size of ultrasound images was cropped to 256 × 256 pixels with strides of 180 and 248 pixels in width and height, respectively.

The implemented GANs are as follows: 
Pix2Pix~\cite{isola2017image}, CycleGAN~\cite{zhu2017unpaired}, UNIT~\cite{liu2017unsupervised}, SRGAN~\cite{ledig2017photo}, TraVeLGAN~\cite{amodio2019travelgan}, NICE-GAN~\cite{chen2020reusing}, and U-GAT-IT~\cite{kim2019u}.
We used Pytorch~\cite{paszke2017automatic} framework to implement and train GANs.
An Adam optimizer with hyper-parameters of $\beta_{1}=0.9$, $\beta_{2}=0.999$, and $\epsilon=10^{-8}$ was used for every models.
The initial learning rate was $10^{-4}$ and divided by half in every 30 epochs.
The models were trained for 120 epochs.
For fine-tuning the process, the learning rate was $10^{-5}$.

For SRGAN, we changed the training scheme from the SR task to the image-to-image translation task.
Since the downsampling process was removed, we removed pixel shuffler blocks for upsampling layers.
We also used a PatchGAN instead of fully connected layers with a sigmoid activation function for the discriminator network.
The authors of~\cite{ledig2017photo} proposed pre-trained VGG loss for improving image super-resolution performance; however, this approach was not suitable for our training scheme.
Therefore, we only used a mean squared error (MSE) loss and a $L_{1}$ loss for the modified SRGAN.
We refer to the modified SRGAN as SRGANUS to avoid confusion.
Except for the changes mentioned above, every training scheme and structure of each network were the same as each network proposed by the respective author.

We performed a quantitative analysis by using the structural similarity index (SSIM)~\cite{wang2004image}, peak signal-to-noise ratio (PSNR), and a root mean square error (RMSE).
We utilized a high-frequency image as a reference image.
The related equations are as follows:

\begin{figure}[t]
\centering
\includegraphics[width=\columnwidth]{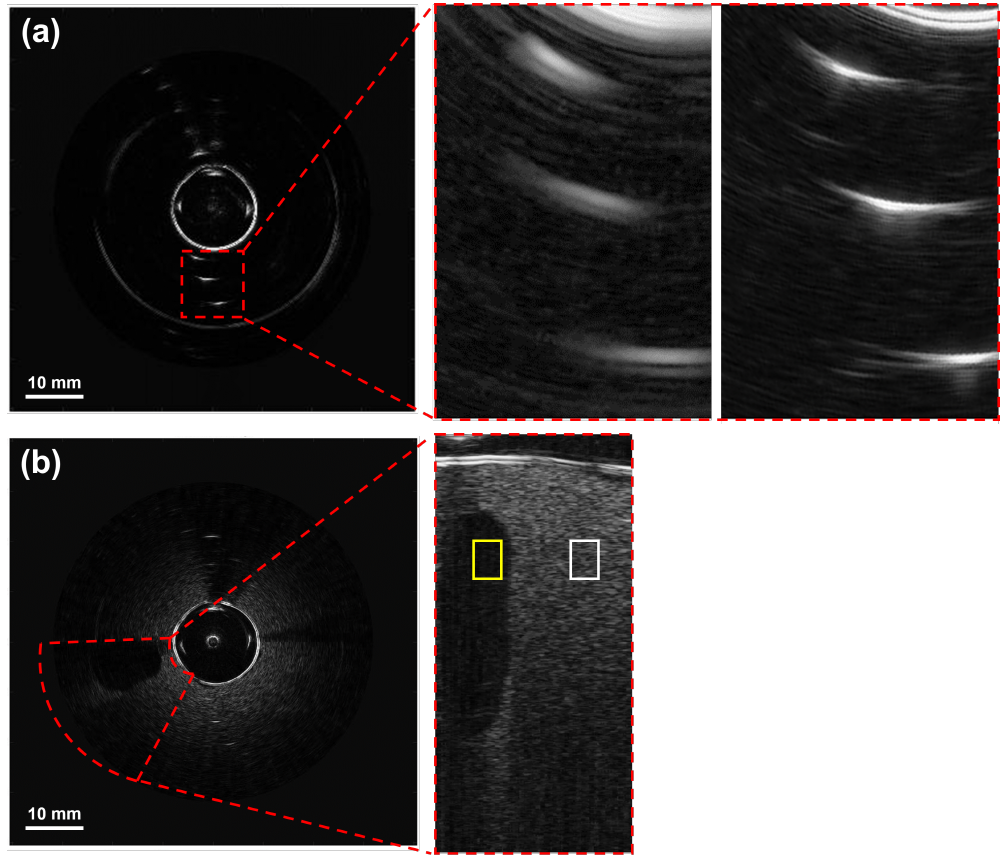}
\caption{Extracted ultrasound images for evaluation of the proposed EUS system. (a) A wire phantom image. The left figure represents a low-frequency ultrasound image of 100 $\mu$m wires, whereas the right figure represents a high-frequency ultrasound image of 100 $\mu$m wires. (b) A tissue-mimicking phantom image. The homogeneous region is highlighted with the yellow solid box, while the speckled region is highlighted with the white solid box.}
\label{fig:6}
\end{figure}

\begin{equation}SSIM(x,y)=\frac{(2\mu_{x}\mu_{y}+C_{1})(2\sigma_{xy}+C_{2})}{({\mu_{x}}^2+{\mu_{y}}^2+C_{1})({\sigma_{x}}^2+{\sigma_{y}}^2+C_{2})},\label{eq:4}\end{equation}

where $\mu_x$ and $\mu_y$ represent the average values of the images and $\sigma_x$, $\sigma_y$, and $\sigma_xy$ represent the standard deviation and cross-covariance of the images. 
We used $(0.01\times L)^2$ and $(0.03\times L)^2$ for $C_1$ and $C_2$, respectively, where the L is the dynamic range.

\begin{equation}MSE=\frac{1}{N} \sum_{i = 1}^{N} ||x_{i} - y_{i}||^2,\label{eq:5}\end{equation}

where $x_{i}$ and $y_{i}$ are the generated image and the reference image, and $N$ represents the number of images.

\begin{equation}PSNR=20\log_{10} (\frac{\textrm{max}(image)}{\sqrt{MSE}}),\label{eq:6}\end{equation}

where $\textrm{max}(image)$ is the maximum intensity of the ultrasound image.

\begin{equation}RMSE=\sqrt{MSE},\label{eq:7}\end{equation}

We also trained and tested an EDSR~\cite{lim2017enhanced} network to investigate the feasibility of the SR task for ultrasound imaging.
We then quantitatively and qualitatively compared results from the SR task and the image-to-image translation task in ultrasound imaging.
All training and testing environments were operated with an Intel i7-7800X CPU (3.50 GHz), 64 GB of RAM, and an NVIDIA RTX 3090 (24 GB) GPU.
The total training process took approximately 3.6 weeks.

\section{RESULTS}
\subsection{Evaluation of the Proposed Dual-element EUS System}

\begin{table}[t]
\centering
\caption{System Evaluation Results. $1^{st}$, $2^{nd}$, and $3^{rd}$ represent the first, second, and third positions of 100 $\mu$m wires.}
\setlength{\tabcolsep}{3pt}
\begin{tabular}{ m{15pt} m{30pt} m{30pt} m{12pt} m{20pt} m{20pt} m{50pt} }
\hline
& 
Lateral ($\mu$m)&
Axial ($\mu$m)&
CNR&
SSNR ($\mu \pm \sigma$)$^{\mathrm{a}}$&
\multicolumn{2}{c}{eSNR (dB)} \\
\hline
&
&
&
&
&
$1^{st}$&
32.610 \\
\cline{6-7}
Low&
2874.278&
606.200&
1.689&
2.651$\pm$0.496&
$2^{nd}$&
31.028 \\
\cline{6-7}
&
&
&
&
&
$3^{rd}$&
\textbf{31.529} \\
\hline
&
&
&
&
&
$1^{st}$&
30.421 \\
\cline{6-7}
High&
\textbf{2630.535}&
\textbf{336.600}&
\textbf{3.030}&
\textbf{4.192$\pm$0.344}&
$2^{nd}$&
32.157 \\
\cline{6-7}
&
&
&
&
&
$3^{rd}$&
28.837 \\
\hline
&
&
&
&
\multicolumn{3}{c}{$^{\mathrm{a}}$mean $\pm$ standard deviation.}
\end{tabular}
\label{tab:1}
\end{table}

Fig.~\ref{fig:6} illustrates the extracted regions where the evaluation of the proposed system with the dual-element EUS probe has proceeded.
We use metrics of the lateral and axial resolution for evaluating the spatial resolution, CNR and SSNR for assessing the image quality, and eSNR for evaluating the imaging depth.

Since the higher-frequency ultrasound transducer can provide a higher spatial resolution, both the lateral and axial resolution show better results than those from the low-frequency transducer, as presented in Table~\ref{tab:1}.
In addition, the results of CNR and SSNR show that the high-frequency ultrasound produces a higher quality image compared to the low-frequency ultrasound image.
The high-frequency image was 1.341 and 1.541 higher than the low-frequency image in terms of CNR and SSNR, respectively.

\begin{figure*}[t]
\centering
\includegraphics[width=1.7\columnwidth]{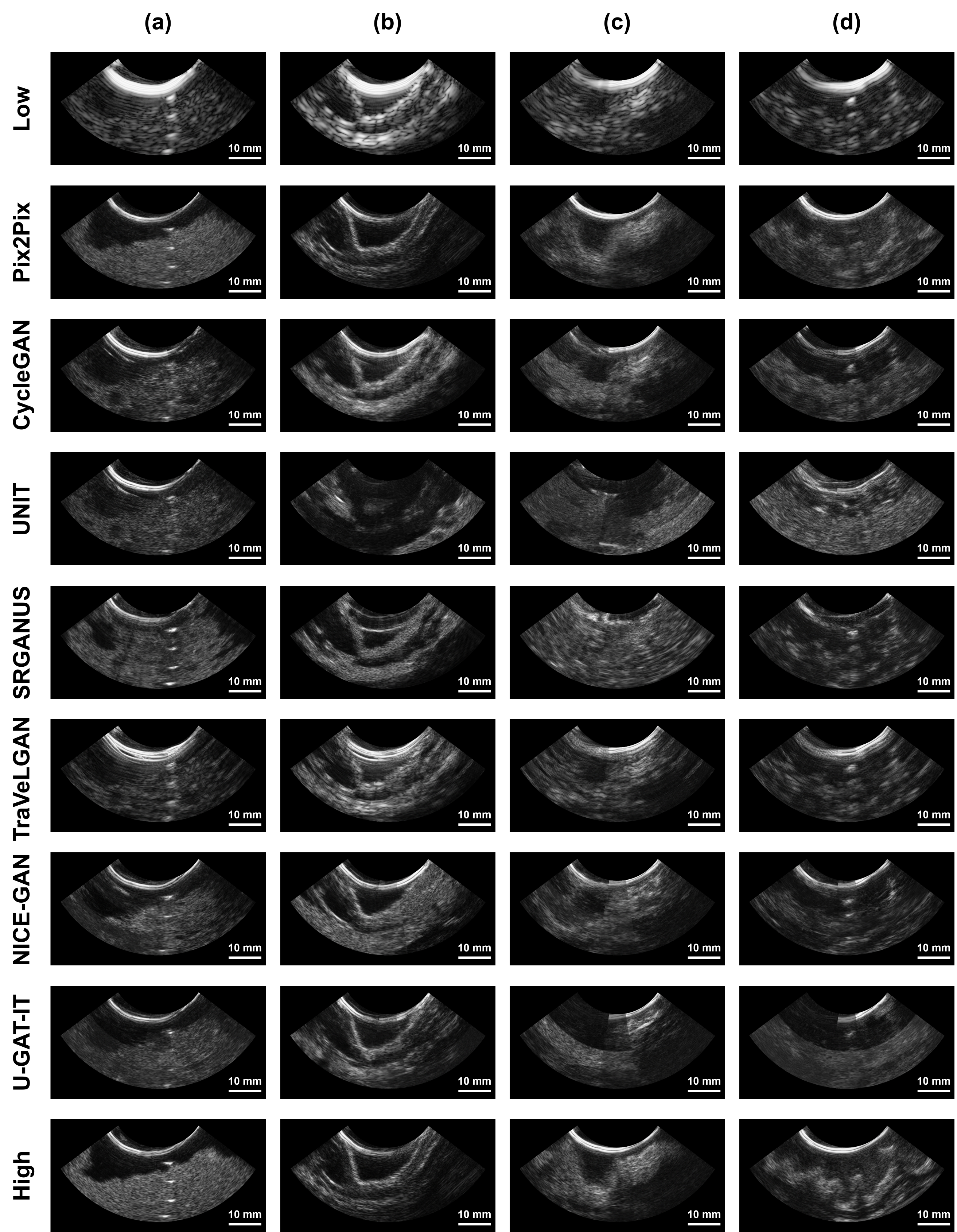}
\caption{Qualitative results for the proposed EUS probe with various deep learning models. (a) A tissue-mimicking phantom. (b) An esophagus. (c) A stomach. (d) An intestine.}
\label{fig:7}
\end{figure*}

In Table~\ref{tab:1}, the listed eSNR values from each wire target show the performance of the penetration depth according to each transducer.
In the high-frequency image, the second wire target exhibits the highest eSNR because the second wire target is located near the focal depth of the high-frequency ultrasound transducer. 
The high-frequency ultrasound image is reduced by 1.584 dB at the third target compared to the first target, while it is reduced by 1.081 dB in the low-frequency image.

\subsection{Quantitative and Qualitative Comparisons for Generated Ultrasound Images}

\begin{figure*}[t]
\centering
\includegraphics[width=1.5\columnwidth]{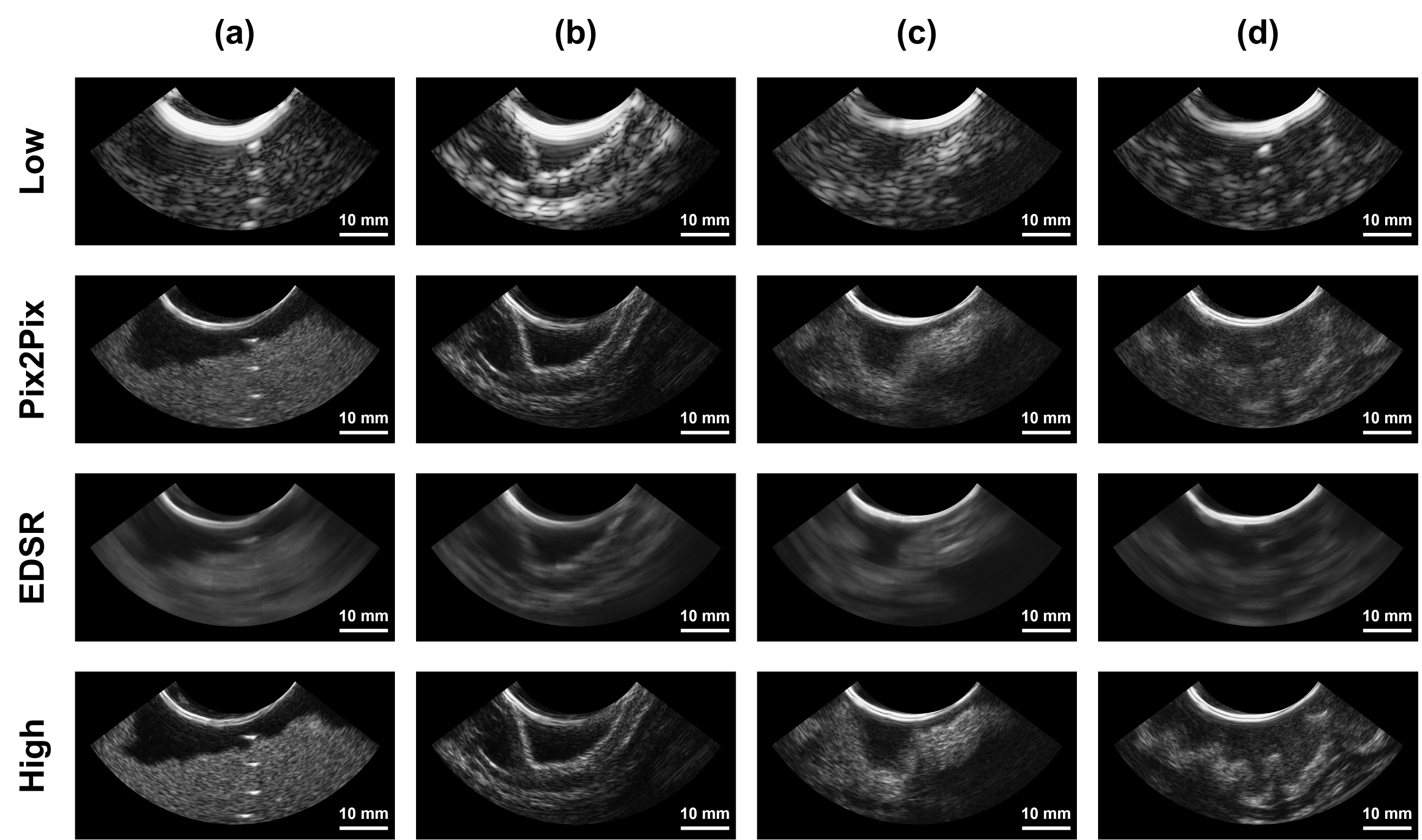}
\caption{Qualitative comparisons of results by the representative GAN and SR model. Low-frequency, Pix2Pix generated, EDSR generated, and reference high-frequency images of (a) A tissue-mimicking phantom. (b) An esophagus. (c) A stomach. (d) An intestine.}
\label{fig:8}
\end{figure*}

Table~\ref{tab:2} shows quantitative results for the GANs and the representative SR model.
Among the GANs, Pix2Pix has the highest SSIM and PSNR.
Interestingly, EDSR, which represents the SR model, demonstrates comparable results with Pix2Pix and even shows better performance in terms of SSIM.

To carry out a pixel-wise image quality assessment, we compared RMSE to each deep learning model.
Pix2Pix shows the lowest RMSE compared to the other networks and its RMSE is 3.4563 less than EDSR.

Fig.~\ref{fig:7} illustrates a qualitative comparison among the GANs.
Similar to the results in Table~\ref{tab:2}, Pix2Pix generates ultrasound images of the tissue-mimicking phantom and various porcine organs that are similar to the high-frequency images.
The other networks are also able to generate speckle patterns from the low-frequency images corresponding to the style of the high-frequency images, as shown in Fig.~\ref{fig:7}.
Similar to the quantitative results, UNIT shows the lowest performance among the other networks.
As shown in Fig.~\ref{fig:7}(c) and (d), Pix2Pix can generate synthetic images similar to the actual high-frequency images while the other networks do not accurately generate the style and structure of the stomach and intestine.
Since we cropped the input images with certain strides, grid artifacts slightly occurred in the overlapping area after the image reconstruction process.
Grid distortion can be overcome by simply upgrading or merging multiple GPUs for training and testing models based on the original image size.

Fig.~\ref{fig:8} shows qualitative results for the representative GAN and SR network.
The results of the image-to-image translation task exhibit detailed structures and speckle information similar to high-frequency ultrasound images.
However, when the scheme of the SR task was applied, EDSR did not generate clear ultrasound images, contrary to the quantitative results in Table~\ref{tab:2}.

\begin{table}[t]
\centering
\caption{Quantitative comparisons for the different GANs and the representative SR model. \textbf{Bold} represents the first-best performance and \underline{Underline} represents the second-best performance.}
\begin{tabular}{ c c c c }
\hline
& 
Mean SSIM&
Mean PSNR (dB)&
Mean RMSE \\
\hline
Pix2Pix~\cite{isola2017image}&
\underline{0.4777}&
\textbf{20.3151}&
\textbf{25.4314} \\
\hline
CycleGAN~\cite{zhu2017unpaired}&
0.3909&
15.4732&
43.2707 \\
\hline
UNIT~\cite{liu2017unsupervised}&
0.3284&
13.4917&
55.8447 \\
\hline
SRGANUS~\cite{ledig2017photo}&
0.3920&
15.0985&
45.3081 \\
\hline
TraVeLGAN~\cite{amodio2019travelgan}&
0.3862&
14.7441&
47.1854 \\
\hline
NICE-GAN~\cite{chen2020reusing}&
0.4066&
16.0786&
40.4237 \\
\hline
U-GAT-IT~\cite{kim2019u}&
0.4178&
15.7671&
42.3021 \\
\hline
\hline
EDSR~\cite{lim2017enhanced}&
\textbf{0.5620}&
\underline{19.0406}&
\underline{28.8877} \\
\hline
\end{tabular}
\label{tab:2}
\end{table}

\section{DISCUSSION}
These results showed the feasibility of synthetic high-resolution in-depth imaging by using the dual-element EUS probe.
Since we reconstructed each high-frequency ultrasound signal with a phase delay of 180° to each low-frequency ultrasound signal according to the PRF, low- and high-frequency ultrasound images are aligned.
System evaluation results demonstrated that our proposed dual-element EUS probe is suitable for acquiring low- and high-frequency ultrasound image pairs for biological tissues.

Because our system is capable of producing a paired image dataset, we investigated the capability of the system for generating high-resolution ultrasound images from low-resolution images using deep learning techniques.
We implemented image-to-image translation networks and the representative SR network.
For image-to-image translation networks, we performed experiments to investigate the most promising network for clinical applications in the paired image dataset.
In supervised or unsupervised manners, we utilized Pix2Pix for the paired image dataset and other GANs specialized for unpaired in two or multi-domain problems.
In particular, Pix2Pix generated both detailed synthetic structures and patterns similar to those from high-frequency images.
On the other hand, the other GANs failed to generate detailed structures with precise speckle patterns.
Since synthetic images for clinical applications should be generated in a direction similar to real high-frequency images as much as possible, deep learning networks should be trained in a direction that causes overfitting.
Therefore, for clinical purposes, it is better to configure a deep learning network in two-domain problems with supervised image-to-image translation manners.

We also trained and tested SRGAN for our application in the same structure and training scheme proposed in a previous study.
However, in the case of the original SRGAN, this network converged to a local minimum and it was impossible to generate valid synthetic high-resolution ultrasound images.
This may be because the original structure and loss functions in SRGAN are specialized for more general SR tasks.
Therefore, we revised SRGAN to SRGANUS for our application and eventually could acquire valid synthetic ultrasound images.

In an SR training manner, we implemented EDSR, which is one of the representative models among SR networks.
As shown in Fig.~\ref{fig:8}, the overall shape and intensity distributions of generated images from EDSR are similar to the high-frequency images, but the detailed structure and speckle patterns are lost.
Therefore, although EDSR showed the highest SSIM and the second-best performance in terms of PSNR, Pix2Pix optimized for the paired image dataset shows better RMSE than EDSR.
Note that Pix2Pix showed the highest PSNR and the second-highest SSIM.

Since the principle of the GAN is to produce a similar probability distribution to real images, random noises are generated depending on the architecture or training scheme of each model.
Therefore, the generated structure or speckle patterns are naturally different compared to those from actual high-frequency images in a pixel level.
These may lead to SSIM from Pix2Pix and quantitative results from the other GANs being far lower than those from EDSR.

From a clinical point of view, the novel dual-element EUS of low frequency and high frequency will be able to produce both low- and high-frequency images almost at the same time, although they have a phase delay of 180° to each low-frequency ultrasound signal. 
This unique feature of the novel dual-element EUS will help endoscopists examine diseases of the GI tract conveniently, compared to the conventional EUS with a switch to change the low or high frequencies, which cannot provide low- and high-frequency images simultaneously. 
In other words, the novel dual-element EUS will provide images of a deep region that is not visible in high-frequency images at almost the same time as in the form of low-frequency images. 
Furthermore, the synthetic high-resolution in-depth imaging technique can provide high-resolution images with high frequency even in deep areas that high-frequency cannot reach.

These two novel techniques will be very useful in clinical practices. 
They will not only save endoscopists the effort of repeatedly pressing switches to change frequencies but also offer the advantage of being able to examine the GI tract in high resolution regardless of the depth.

Taking a 5 cm gastric subepithelial tumor as an example, the near margin of the tumor from the conventional EUS can be observed relatively well, but it is difficult to observe the far margin of the tumor with the conventional EUS, making it difficult to determine whether the tumor invades other organs. 
In the case of a tumor of the pancreas, the area close to the conventional EUS can be observed in high quality, but it is difficult to obtain high-definition images of the tumor area far from the EUS. 
As another example of a tumor of the pancreas, when performing an ultrasound-guided fine-needle biopsy, if the tumor is far from the EUS, it may not be possible to obtain a high-quality image of the tumor, and thus it may not be possible to observe the blood vessels inside or around the tumor properly. 
In this case, there is a risk that when the tumor is punctured with a fine needle for biopsy, it may puncture the blood vessels and cause serious bleeding.
Therefore, the novel dual-element EUS of low and high frequency and synthetic high-resolution in-depth imaging technique will be very useful for endoscopists.

\section{CONCLUSION}
We reported a novel approach to overcome the inherent limitations of endoscopic ultrasound imaging.
We developed a dual-element EUS probe with a distinctive scanning scheme to enable the same image plane for both low- and high-frequency ultrasound transducers.
We also examined the feasibility of enhancing clinical ultrasound images on several deep learning networks with a paired image dataset.
It was found that the conventional SR models were not suitable for the generation of high-resolution images from a low-resolution image, whereas the image-to-image translation models optimized in two-domain problems in a supervised manner using the paired image dataset are better suited for this.
Fine-tuning of the constructed networks with low- and high-resolution image pairs, obtained by using a dual-element EUS probe, made it possible to obtain better synthetic high-resolution in-depth ultrasound images. 
The synthetic high-resolution in-depth ultrasound images may facilitate practical diagnosis of examining the GI tract for clinicians.
In future work, with a slight modification of the probe structures, we will further improve a versatile attachable EUS probe for high-resolution in-depth ultrasound imaging.

\bibliographystyle{IEEEtran}
\bibliography{tmi}

\end{document}